\documentclass{ws-ijmpa}

\begin{document}

\markboth{A. Szczurek and P. Paw{\l}owski}
{$\pi^+ - \pi^-$ asymmetry}

%
\catchline{}{}{}{}{}
%

\title{$\pi^+ - \pi^-$ ASYMMETRY \\ AND THE NEUTRON SKIN IN HEAVY NUCLEI}

\author{\footnotesize ANTONI SZCZUREK}

\address{Institute of Nuclear Physics PAN, ul. Radzikowskiego 152\\
PL-31-342 Cracow, Poland;\\
University of Rzesz\'ow, ul. Rejtana 16c\\
PL-35-959 Rzesz\'ow, Poland
}

\author{PIOTR PAW{\L}OWSKI}

\address{Institute of Nuclear Physics PAN, ul. Radzikowskiego 152\\
PL-31-342 Cracow, Poland
}

\maketitle

\pub{Received (Day Month Year)}{Revised (Day Month Year)}

\begin{abstract}
In heavy nuclei the spatial distribution
of protons and neutrons is different.
At CERN SPS energies production of $\pi^+$ and $\pi^-$
differs for $pp$, $pn$, $np$ and $nn$ scattering.
These two facts lead to an impact parameter dependence
of the $\pi^+$ to $\pi^-$ ratio in $^{208}Pb + ^{208}Pb$ collisions.
A recent experiment at CERN seems to confirm qualitatively these
predictions.
It may open a possibility for determination of neutron density
distribution in nuclei.

\end{abstract}

\section{Introduction}

Electron scattering off nuclei provides direct information about
charge distribution \cite{Frois77}, which is
closely related to the spatial distribution of protons.
The information about the neutron spatial distribution is not
accessible directly.

The radiochemical method applied to
antiprotonic atoms \cite{radiochemical3}
allows one to measure the neutron-to-proton ratio at the peripheries
of nuclei.
Another method is based on the analysis of X-ray spectra of
antiprotonic atoms \cite{Trzcinska}.

On the theoretical side the difference between the proton and
neutron distributions can be obtained in the framework
of Hartree-Fock (HF) method (see for example \cite{HL98})
or Hartree-Fock-Bogoliubov (HFB) method (see for example
\cite{HFB_nuclear_skins}).

Stable heavy nuclei exhibit an excess of neutrons over protons.
A recent experiment at CERN \cite{Rybicki}
for charged pion production in the $^{208}Pb + ^{208}Pb$ collision
has observed an interesting dependence of the ratio
$R_{+/-} = \frac{d \sigma}{d x_F}^{\!\!\pi^+} /
 \frac{d \sigma}{d x_F}^{\!\!\pi^-}$
on the Feynman variable $x_F$.
Surprisingly, the ratio for central and peripheral collisions
differs significantly \cite{Rybicki}.
This presentation is based on \cite{PS04} where more details
are given.

\section{Estimate of the effect}
We assume that pions are produced in elementary
N-N collisions. Then
\begin{equation}
\frac{d \sigma^{A_1 A_2 \rightarrow \pi^{\pm}}}{d x_F}(b,x_F;W) =
\sum_{\alpha \beta = p,n} N_{\alpha \beta \rightarrow \pi^{\pm}}(b;W) \;
\frac{d \sigma^{\alpha \beta \rightarrow \pi^{\pm}}}{d x_F}(x_F;W)
\; ,
\label{NN_scattering}
\end{equation}
where $W$ is energy per binary N-N collision and
$N_{\alpha \beta \rightarrow \pi^{\pm}}$ are numbers
of collisions of a given type.
Only elementary cross sections for the $pp \rightarrow \pi^{\pm}$
processes are known experimentally.
Therefore in the following we shall use 
$\frac{d \sigma}{d x_F} (x_F;W)$ calculated in
the HIJING model \cite{HIJING}.

It is often assumed that the dynamics of nuclear collisions
is governed by the number of binary collisions \cite{binary_collisions}.
The number of binary N-N collisions at a given impact
parameter is proportional to the nucleus-nucleus thickness
\begin{equation}
T_{A_1 A_2}(\vec{b}) =
\int d^2 s_1 T_{A_1}(\vec{s}_1) \; T_{A_2}(\vec{s}_1 - \vec{b}) =
\int d^2 s_2 T_{A_1}(\vec{s}_2 - \vec{b}) \; T_{A_2}(\vec{s}_2) \; .
\label{coll_thick}
\end{equation}
In Eq.(\ref{coll_thick}) we introduced
$T_{A_i}(\vec{b}) = \int d z_i \; \rho_{A_i}(\vec{b},z_i)$,
where $\rho_{A_i}$ is the density function of the nucleus $A_i$
normalized to the number of nucleons.
Analogously the number of binary collisions of a given type
in Eq.(\ref{NN_scattering}) can be written as
\begin{equation}
N_{\alpha \beta \rightarrow \pi^{\pm}}(b; W) =
T_{A_1 A_2}^{\alpha \beta}(\vec{b}) \cdot \sigma_{ine}^{NN}(W) \; ,
\label{fraction_binary}
\end{equation}
where
\begin{equation}
T_{A_1 A_2}^{\alpha \beta}(\vec{b}) =
\int d^2 s_1 T_{A_1}^{\alpha}(\vec{s}_1)
          \; T_{A_2}^{\beta}(\vec{b}-\vec{s}_1) =
\int d^2 s_2 T_{A_1}^{\alpha}(\vec{b}-\vec{s}_2)
          \; T_{A_2}^{\beta}(\vec{s}_2)
\; ,
\label{generalized_coll_thick}
\end{equation}
where now $T_{A_i}^p$ and $T_{A_i}^n$ are nucleus thicknesses of
protons and neutrons, respectively.
We have assumed one universal inelastic cross section
$\sigma_{ine}^{\alpha \beta}(W) = \sigma_{ine}^{NN}(W)$.

As a second limiting case we consider the wounded nucleon model
\cite{wounded_nucleon}. We assume that the production
of particles is proportional to the number of wounded nucleons.
Then the cross section for the nuclear collision can be written as
\begin{eqnarray}
 \frac{d\sigma ^{A_{1}A_{2}\rightarrow \pi ^{\pm}}}{dx_{F}}(b,x_{F};W)
\propto \sum _{\alpha ,\beta =p,n}
\left[N_{\alpha /A_{1}}^{wou}(b,W)\; w_{2}^{\beta }(b)\;
 \frac{d\sigma ^{\alpha \beta \rightarrow \pi ^{\pm }}}{dx_{F}}(x_{F};W)\right.
 \nonumber
   \\
+\left. N_{\alpha /A_{2}}^{wou}(b,W)\; w_{1}^{\beta }(b)\;
  \frac{d\sigma ^{\beta \alpha \rightarrow \pi ^{\pm
      }}}{dx_{F}}(x_{F};W)\right]  \; .  
 \label{wounded_nucleon_model}
\end{eqnarray}
In the formula above $N_{\alpha/A_i}^{wou}$ is the number of
wounded $\alpha$ (p or n) in nucleus $A_1$ or $A_2$
and $w_i^{\beta}$ is the probability that the wounded $\alpha$
interacted with $\beta$ (p or n) from nucleus $A_2$ or $A_1$,
respectively.
Eq.(\ref{wounded_nucleon_model}) is equivalent to
Eq.(\ref{NN_scattering})
with
\begin{equation}
N_{\alpha \beta \rightarrow \pi^{\pm}} =
      N_{\alpha/A_1}^{wou}(b,W) \; w_2^{\beta}(b)
    + N_{\beta/A_2}^{wou}(b,W) \; w_1^{\alpha}(b) \; .
\label{fractions_wounded}
\end{equation}

By construction, in our approach the numbers of wounded protons
and neutrons reproduce the well known formula from \cite{wounded_nucleon}
for the number of wounded nucleons
\begin{equation}
N_{N/A_i}^{wou}(b) = N_{p/A_i}^{wou}(b) + N_{n/A_i}^{wou}(b) \; .
\label{N_wounded_sum}
\end{equation}
Our construction requires also
\begin{equation}
w_i^p(b) + w_i^n(b) = 1 \; .
\label{N_wounded_fraction_sum}
\end{equation}
The fractions $w_i^{\alpha}$ were estimated as
\begin{equation}
w_i^{\alpha}(b) = \frac{N_{\alpha/A_i}^{wou}(b)}
{N_{p/A_i}^{wou}(b) + N_{n/A_i}^{wou}(b)} \; ,
\label{w_i}
\end{equation}
which by construction fulfils (\ref{N_wounded_fraction_sum}).
We use proton and neutron densities
calculated in the HFB method \cite{HFB_nuclear_skins}
with Skyrme interaction SLy4.

\begin{figure}[htb] 
\begin{center}
\includegraphics[width=7cm]{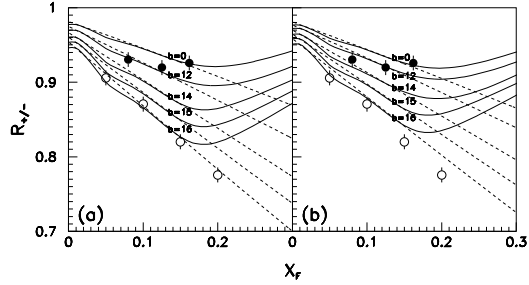}
\end{center}
\caption{
The ratio $R_{+/-}$ as a function of $x_F$
for the model with binary collision (left panel)
and for the model with the scaling with number of wounded
nucleons (right panel) for selected values of impact parameters.
Elementary cross sections are taken from HIJING model (solid lines) or
from our fit to NA49 experimental $p+p$ and $p+n$ data  (dashed lines).
\label{fig_R_xf}
}
\end{figure}
In Fig.\ref{fig_R_xf} we present the ratio $R_{+/-}$
as a function of $x_F$ for different values of $b$
in the two models considered.
For comparison we show
preliminary experimental data for "central collisions" (solid circles)
and "peripheral collisions" (open circles) from \cite{Rybicki}.
The notion of central and peripheral collisions was not specified
in \cite{Rybicki}. Therefore the data can be used only as an indication
of the effect. 
Our approach explains the experimental data provided they are
extremely peripheral.

\begin{figure}[htb] 
\begin{center}
\includegraphics[width=7cm]{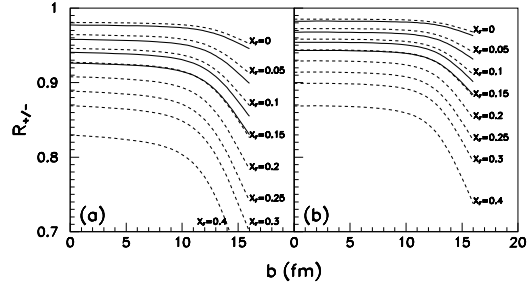}
\end{center}
\caption{
The ratio $R_{+/-}$ as a function of impact parameter
for the model with binary collision (left panel)
and for the model with the scaling with number of wounded
nucleons (right panel) for selected values of $x_F$.
\label{fig_R_b}
}
\end{figure}

Our results are reliable up to $x_F \approx$
0.2. It is known that the HIJING code does not describe the
$\pi^+$-$\pi^-$ asymmetry in elementary collisions in the region
of large $x_F$. In principle, instead of model calculations of
the elementary cross sections one could use directly experimental
data.
Using isospin symmetry and aproximate relations
between elementary cross sections \cite{PS04}
one can write for $x_F>$0.1:
\begin{equation}
R_{+/-}(x_F) = \frac{N_{pp}(b)+N_{pn}+r(x_F) [N_{np}(b)+N_{nn}(b)]}
{r(x_F)[N_{pp}(b)+N_{pn}(b)]+N_{np}(b)+N_{nn}(b)} \; ,
\label{experimental_R}
\end{equation}
where
\begin{equation}
r(x_F) \equiv \frac{d \sigma^{pp \to \pi^-}}{d \sigma^{pp \to \pi^+}} \; .
\label{exp_aux}
\end{equation}
We fit the NA49 data shown in Fig.2a of Ref.\cite{Rybicki} with
the simple form $r(x_F) = c_r (1-x_F)^{n_r}$. The corresponding
results for $R_{+/-}(x_F)$ are shown in Fig.\ref{fig_R_xf}
by the dashed lines. We find a good agreement with the calculations based
on HIJING below $x_F \approx$ 0.2. Above $x_F \approx$ 0.2 the
results based directly on experimental data lay below those based
on HIJING.

For completeness in Fig.\ref{fig_R_b} we present the ratio
$R_{+/-}$ as a function of the impact parameter $b$ for different
values of $x_F$.

\section{Summary}

Surprisingly the binary collision picture gives very similar results
to the predictions of the wounded nucleon model.
This suggests that a detailed comparison of model results
with the well defined ($x_F$, $b$) experimental data
could open a new possibility to study the neutron density
profile. We expect that the NA49 collaboration
at CERN will be able to gather the corresponding experimental data
in the near future. Can it provide a method competitive
to that offered by proton-nucleus elastic scattering, antiprotonic atoms
or parity violating electron scattering? Of course results of
these methods must finally converge. Therefore one may hope that
together they will provide more reliable information on
neutron distribution in nuclei.


\end{document}